\documentstyle[prc,aps,epsf,amssymb,amsbsy,amstext,subeqnarray,amsfonts]{revtex}
\begin{document}

\title{Comment about pion electro-production and the axial form factors}

\author{P.A.M. Guichon}
\address{{SPhN/DAPNIA, CEA-Saclay, F91191 Gif
sur Yvette Cedex} }

\maketitle

\begin{abstract}
The claim by Haberzettl\cite{ducon} that the axial form
factor of the nucleon cannot be accessed through threshold pion electroproduction is unfounded.
\end{abstract}
\vskip 10pt
The soft pion theorem\cite{adler} for pion electro-production off a nucleon has
the form:
\begin{equation}
\label{Eq1}
f_{\pi }M(\gamma ^{*}+N\rightarrow \pi (q_{\mu}=0)+N)=-i\lim _{q\rightarrow 0}\left( q_{\mu }\langle N|
 \int d^{4}xe^{iq.x}T(A^{\mu }(x),\varepsilon .J(0))| N\rangle \right) 
 -\left\langle N\left| [Q_{5},J.\varepsilon ]\right| N\right\rangle .
\end{equation}
The commutator \( [Q_{5},J.\varepsilon ] \) is given by current algebra and
is equal to the axial current. So, up to corrections of order \( m_{\pi }\), 
this reaction gives access to the axial form factors\cite{remarque} of the nucleon. One can also derive (\ref{Eq1})
{}``à la Adler{}'' using PCAC in the presence of an electro-magnetic field.

The first term on the RHS of (\ref{Eq1}) is the amplitude \( T_{A\gamma } \)
for producing an axial current \( A^{\mu } \) by the electro-magnetic
interaction \( \varepsilon .J(0) \). The crucial point
is that, because of the factor \( q_{\mu } \), only the part of \( T_{A\gamma } \)
which is singular at \( q_{\mu }=0 \) can contribute in the limit \( q_{\mu }\rightarrow 0. \)
This implies that the only diagrams which survive are those where the axial
current is attached to an \emph{external} leg, and in the present case the only
external legs are those of the nucleon. Therefore in Fig.3 of Ref.\cite{ducon}
only the first two diagrams have to be kept because \emph{the sum of the others
vanishes when one contracts with \( q_{\mu } \) and takes the soft pion limit.} 

The confusing point in Ref.\cite{ducon} is that the author has split the axial
current in what he calls a {}``weak{}'' part \( J_{A,W} \) and a {}``hadronic{}''
part \( J_{A,H} \), so that only \( g_{A}(t) \) appears in the {}``weak{}''
part. The price to pay for this strange splitting is the presence of an unphysical
pole at \( t=0 \) in both \( J_{A,W} \) and \( J_{A,H} \). Of course these
poles cancel out in the full current. The trap in the reasoning of Ref.\cite{ducon}
is that the contributions of \( J_{A,W} \) and \( J_{A,H} \) to \( T_{A\gamma } \)
are calculated separately. Due to the unphysical pole the author finds a finite
contribution due to \( J_{A,H} \) but he leaves the contribution due to \( J_{A,W} \)
unspecified, arguing that it cannot be computed explicitly due to the nucleon
structure and that all what matters is that it depends only on \( g_{A}(t). \)
This is the basis for his argument and it is of course completely misleading
since we do know that the pole part of \( J_{A,W} \) must cancel exactly the
one of \( J_{A,H} \), and the rest vanishes in the soft pion limit. In others
words the quantity called \( \cal W \) by the author is actually zero in the
soft pion limit, independently of the nucleon structure. This is enough to invalidate
the conclusions of Ref.\cite{ducon}.

\vfill{SPhN-00-76}
\end{document}